\documentclass[aps, prx, reprint, lengthcheck, footinbib, showpacs, superscriptaddress, citeautoscript, amsmath, amssymb]{revtex4-2}

\usepackage[utf8x]{inputenc}

\usepackage{amsmath}
\usepackage{graphicx}
\usepackage{float}
\usepackage{textgreek}
\usepackage{chemgreek}
\usepackage{mhchem}
\usepackage{subfigure}
\usepackage{subcaption}
\usepackage{bm} %bold math 
\usepackage{dcolumn} % Align table columns on decimal point
\usepackage[colorlinks=true, citecolor=blue, linkcolor=blue]{hyperref}%
\usepackage{xcolor}

\allowdisplaybreaks

\draft % marks overfull lines with a black rule on the right

\begin{document}
\title{Nonparabolic dispersion of charge carriers in CsPbI$_3$ in the orthorhombic phase.}
\author{O. S. Sultanov}
\affiliation{Spin Optics Laboratory, St. Petersburg State University, Ulyanovskaya 1, Peterhof, St. Petersburg, 198504, Russia}
\affiliation{Faculty of Physics, St. Petersburg State University, Ulyanovskaya 1, Peterhof, St. Petersburg, 198504, Russia}
\author{D. K. Loginov}
\affiliation{Spin Optics Laboratory, St. Petersburg State University, Ulyanovskaya 1, Peterhof, St. Petersburg, 198504, Russia}
\author{I. V. Ignatiev}
\affiliation{Spin Optics Laboratory, St. Petersburg State University, Ulyanovskaya 1, Peterhof, St. Petersburg, 198504, Russia}
\affiliation{Faculty of Physics, St. Petersburg State University, Ulyanovskaya 1, Peterhof, St. Petersburg, 198504, Russia}
\author{D. V. Pankin}
\affiliation{Center for Optical and Laser Materials Research, St. Petersburg State University, Ulyanovskaya 5, St. Petersburg 198504, Russia}
\author{M. B. Smirnov}
\affiliation{Faculty of Physics, St. Petersburg State University, Ulyanovskaya 1, Peterhof, St. Petersburg, 198504, Russia}
\author{M. S. Kuznetsova}
\affiliation{Spin Optics Laboratory, St. Petersburg State University, Ulyanovskaya 1, Peterhof, St. Petersburg, 198504, Russia}
%\tableofcontents

\begin{abstract}

The dispersion curves for the electrons and holes in CsPbI$_3$ in the orthorhombic phase are calculated using the density functional theory (DFT), with the spin-orbit coupling taken into account. The effective masses of the charge carriers are obtained using the parabolic approximation of the dispersion curves in different directions in the $k$-space. It is found that the dispersion curves demonstrate strong nonparabolicity at energies above 0.2~eV for electrons and above 0.1~eV for holes, available for experimental study by the means of optical spectroscopy. We propose a model that describes the dispersion dependences of charge carriers at those energies, where the effective masses of the quasiparticles depend quadratically on the wave vector. An expression is obtained according to the model, which can accurately approximate the dispersion curves for the electron and the hole in all symmetric directions from the center to the boundary of the Brillouin zone.

\end{abstract}

\maketitle %\maketitle must follow title, authors, abstract and \pacs
\section{Introduction}

The optical materials based on lead halide perovskite nanocrystals are promising due to their unique optical \cite{li_cspbx3_2016, yakunin_low-threshold_2015, ramasamy_all-inorganic_2016, akkerman_genesis_2018, nedelcu_fast_2015}, optoelectronic \cite{tan_bright_2014, protesescu_nanocrystals_2015, cho_overcoming_2015, kovalenko_properties_2017, diroll_hightemperature_2017, chen_twodimensional_2017, dey_state_2021, zhang_stable_2024}, and photovoltaic \cite{hodes_perovskite-based_2013, eperon_formamidinium_2014, nie_high-efficiency_2015, zhang_bication_2017, huang_exploration_2018, dey_state_2021} properties. The study of CsPbX$_3$ (X= Cl, Br, I) perovskite nanocrystals (NCs) is a highly challenging problem in optical materials science. Particularly, prospects of application of the CsPbX$_3$-based NCs have been demonstrated for lasers \cite{saouma_selective_2017}, polarizers \cite{wang_polarized_2016}, light-emitting diodes \cite{tan_bright_2014, pathak_perovskite_2015, cho_overcoming_2015, liu_halide-rich_2017}, solar cells \cite{kojima_organometal_2009, lee_efficient_2012, burschka_sequential_2013, hodes_perovskite-based_2013, mei_hole-conductorfree_2014, nie_high-efficiency_2015, zhang_bication_2017, huang_exploration_2018}, and photodetectors \cite{yang_recent_2019}. The knowledge of fundamental properties of these materials is important for the design of practical devices. 

The electronic energy band structure is one of particular importance in this regard and can be studied both experimentally and theoretically \cite{kirstein_lande_2022, kirstein_mode_2023, nestoklon_tailoring_2023}. Experimentally, it has been studied for perovskite NCs through various optical methods, such as the measurement of photoluminescence (PL) and the photoluminescence excitation spectra (PLE) \cite{kolobkova_perovskite_2021, nestoklon_tailoring_2023}. In particular, those methods allow for the observation of not only the ground states of charge carriers (electrons and holes) but also the quantum-confined excited states in various nanostructures. This enables the experimental study of dispersion dependences in the valence and conduction bands. The experimental results show (\cite{kulebyakina_temperature-dependent_2024}) that it is possible to resolve quantum-confined states with energies of a fraction of an electron volt.

However, these experimental studies should be supported by theoretical analysis. The relevance of such theoretical studies is due to the fact that the commonly used effective mass model no longer holds at those energies, and the dispersion curves assume a non-parabolic shape~\cite{ekenberg_nonparabolicity_1989, tomic_influence_2004, PhysRevB.99.085207}. Additionally, no comprehensive theory describing the nonparabolicity of the charge carrier dispersion in perovskites has been proposed so far, despite the importance of the effect on the optical properties of the material.

This paper is dedicated to the development of the theoretical model that describes the nonparabolic dispersion of the charge carriers in lead halide perovskites. We propose an approach based on the DFT calculations of the electronic band structure of CsPbI$_3$ in the orthorhombic phase. The basic principles of such calculations for the lead-halide perovskites and other crystals can be found in Refs.~\cite{nestoklon_tight-binding_2021, kirstein_lande_2022, nestoklon_tailoring_2023, pankin1, pankin2, pankin3, pankin4, pankin5, Sutton2018_CsPbI3_effective_mass}. The data obtained from thes calculations are used to determine the energy range in which the effective mass model for the charge carriers is applicable. Then, a phenomenological model is suggested, which can describe the dispersion dependences of the electrons and the holes with greater precision in a greater range of energies compared to the effective mass model.

\section{Simulation of the crystal band structure of CsPbI$_3$}
\begin{figure}[t]
\centering
\makebox[0.45\columnwidth][c]{%
\subfigure[]{\includegraphics[scale=0.21]{"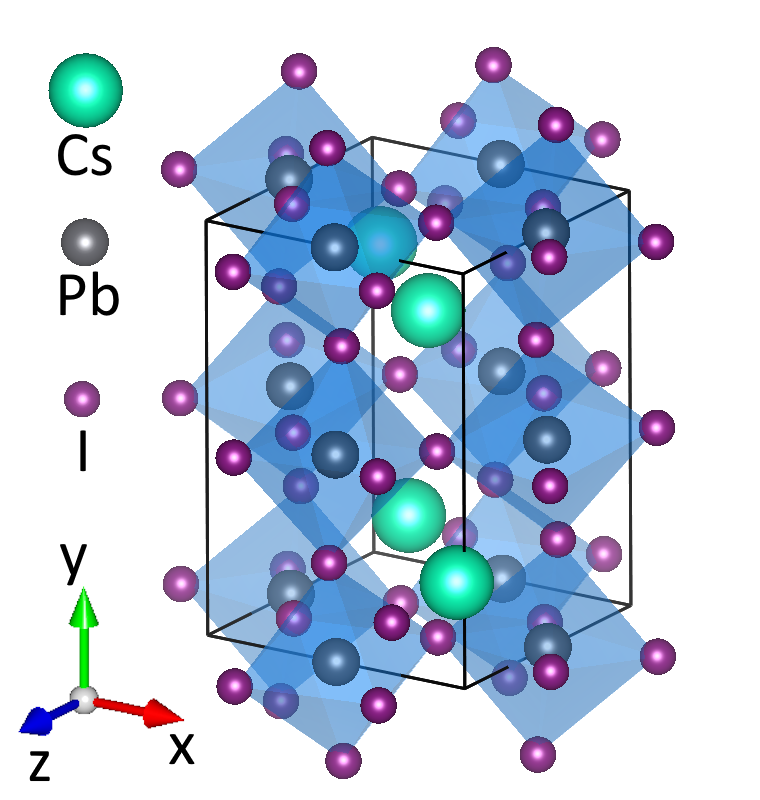"}}
\subfigure[]{\includegraphics[scale=0.21]{"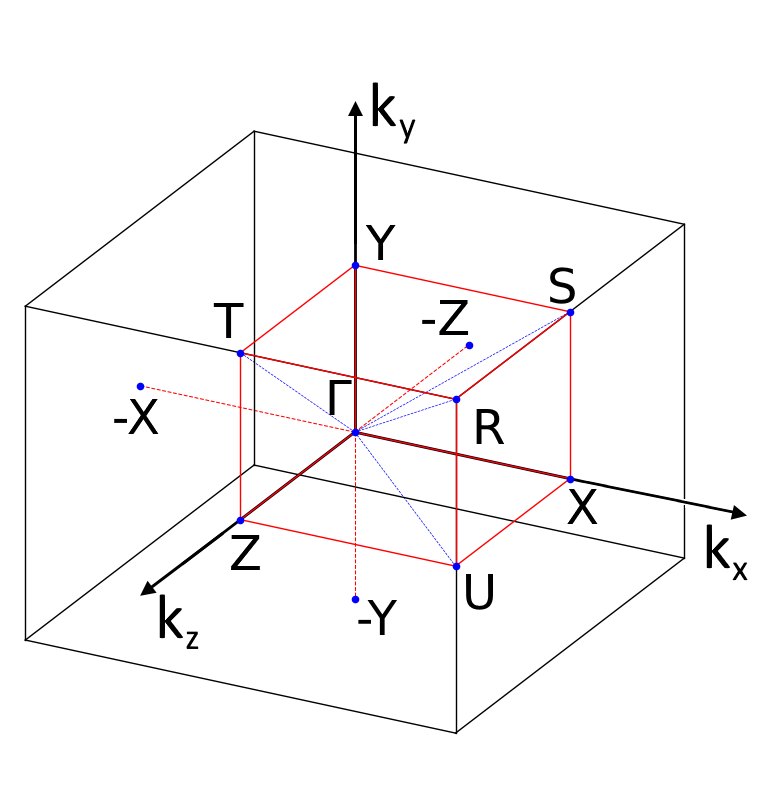"}}
}
\caption{(a) The unit cell of the orthorhombic CsPbI$_3$ crystal lattice. The lattice constants obtained from the DFT calculation are, $a_x = 0.907$ nm, $a_y = 1.254$ nm, and $a_z = 0.801$ nm. The values of the constants agree well with those obtained via X-ray diffraction in~\cite{Sutton2018_CsPbI3_effective_mass}. (b) The first Brillouin zone with the primitive reciprocal lattice vectors, $b_{\alpha} = 1/a_{\alpha}$, where $\alpha=x,y,z$. The labels $\Gamma$, $\mathrm{X}$, $\mathrm{Y}$, etc. show the high-symmetry points used in the DFT calculations. Crystallographic data retrieved from the Materials Project for CsPbI$_3$ (mp-1120768) from database version v2025.09.25. \cite{jain_commentary_2013, horton_accelerated_2025}} 
\label{fig:BrillouinZone}
\end{figure}

It has been shown~\cite{Sutton2018_CsPbI3_effective_mass} that CsPbI$_3$ has an orthorhombic structure at low temperatures, with the respective phase known as $\gamma$-CsPbI$_3$. The respective unit cell, as well as the first Brillouin zone, is schematically shown in Fig. \ref{fig:BrillouinZone}. The unit cell contains four formula units of CsPbI$_3$. The coordinates of the high symmetry points are given in Tab. \ref{tab:ZonePoints}. The wave vector components, $k_x$, $k_y$, and $k_z$, are measured in units proportional to the inverse lattice constants, $a_x$, $a_y$, and $a_z$, respectively.

\begin{table}[t]
\centering
\caption{The relative coordinates of the high symmetry points in the first Brillouin zone in units of the vectors of the reciprocal lattice, $2\pi/a_x = 6.93$ nm$^{-1}$, $2\pi/a_y = 5.01$ nm$^{-1}$, and $2\pi/a_z = 7.84$ nm$^{-1}$.}
\begin{tabular}{|l|l|l|l|}
\hline
Point & $k_x,\frac{2\pi}{a_x}$ & $k_y,\frac{2\pi}{a_y}$ & $k_z,\frac{2\pi}{a_z}$ \\ \hline
Z     & 0.0                    & 0.0                    & 0.5                    \\ \hline
T     & 0.0                   & 0.5                  & 0.5                    \\ \hline
Y     & 0.0                    & 0.5                   & 0.0                    \\ \hline
S     & 0.5                   & 0.5                    & 0.0                    \\ \hline
X     & 0.5                    & 0.0                    & 0.0                    \\ \hline
U     & 0.5                    & 0.0                    & 0.5                    \\ \hline
R     & 0.5                   & 0.5                    & 0.5                    \\ \hline
\end{tabular}
\label{tab:ZonePoints}
\end{table}

In order to perform the calculation of the energy band structure, the generalized gradient approximation (GGA) of density functional theory (DFT) with the Perdew-Burke-Ernzerhoff (PBE) functional is chosen \cite{pankin1}. The Tkatchenko-Scheffler (TS) approach is used to take the dispersion corrections \cite{pankin2} into account. The spin-orbit coupling (SOC) is taken into consideration. The norm-conserving pseudopotentials \cite{pankin1} and the plane wave basis set with a 1200 eV cutoff energy are used. The self-consistent field convergence criterion is $ 10^{-7}~\mathrm{eV/atom}$. The calculations are performed in CASTEP software \cite{pankin1,pankin3}.

At the initial stage, the crystal geometry optimization was performed using the modified Broyden-Fletcher-Goldfarb-Schanno (LBFGS) approach \cite{pankin4,pankin5} until the residual forces, residual stresses, and maximum displacements were smaller than $0.01~\mathrm{eV/\textup{\AA}}$, 0.02 GPa, and $5 \times 10^{-4} \textup{\AA}$, respectively. In the reciprocal space, the Monkhorst-Pack grid of points was used with the k-vector step equal to $0.04~\mathrm{1/\textup{\AA}}$ \cite{pankin6}. The band structure calculations were performed within the 10 eV energy range with a 0.96 eV scissor operator and a $ 0.006~\mathrm{1/\textup{\AA}}$ separation. The electron energy states are calculated for many points in the Brillouin zone to accurately reproduce the dispersion curves for charge carriers.

\begin{figure}[t]
\centering
\includegraphics[width=\columnwidth]{"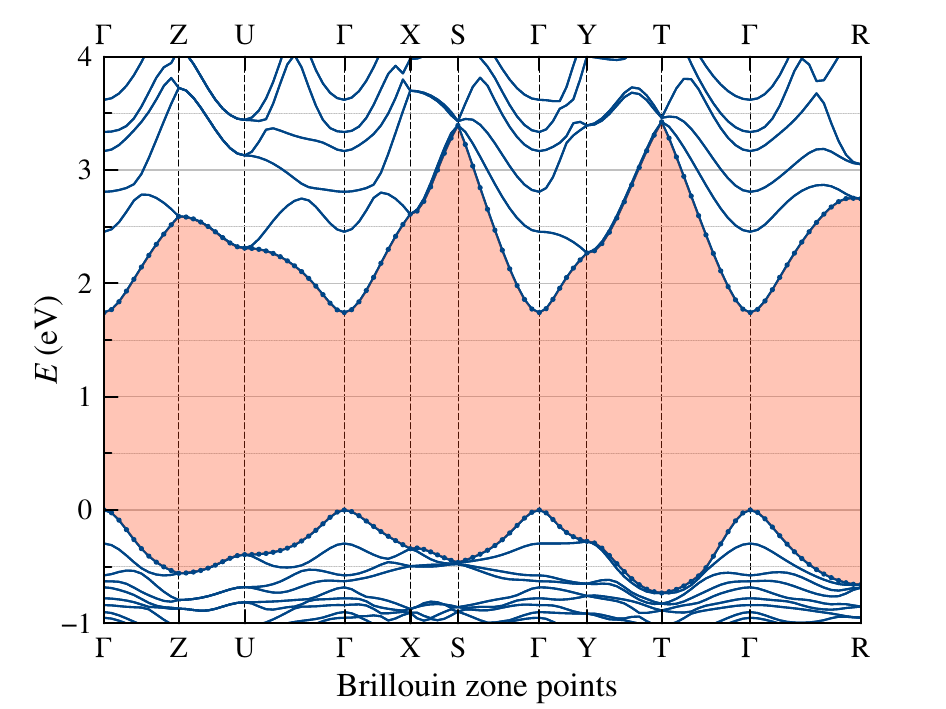"}
\caption{The dispersion dependences for the valence and the conduction bands in $\gamma$-CsPbI$_3$ calculated via the DFT method, with the spin-orbit coupling taken into account.}
\label{fig:DispersionDependences}
\end{figure}

The energy structure of the orthorhombic CsPbI$_3$ crystal calculated via the DFT method is shown in Fig. \ref{fig:DispersionDependences}. The figure shows the dispersion dependences, $E (\vec{k})$, for the electrons in multiple conduction subbands and for the holes in multiple valence subbands. Note that the narrowest band gap of the orthorhombic CsPbI$_3$ crystal is at the $\Gamma$-point. As seen in Fig. \ref{fig:DispersionDependences}, the dispersion curves of different subbands approach each other and become degenerate at critical points of the Brillouin zone. Moreover, some of the dispersion curves undergo anti-crossings and are split due to their interaction. Similar behavior is observed for both the valence and the conduction subbands. 

The accuracy of the obtained DFT results can be verified by comparison with the available experimental data. The DFT calculations well reproduces the lattice constants of the $\gamma$-CsPbI$_3$ obtained via X-ray diffraction~\cite{Sutton2018_CsPbI3_effective_mass}. The band gap energy calculated in the DFT, $E_g=1.7423$ eV, also well agrees with that obtained from the experimental optical spectra in Ref.~\cite{Gau_JLum2023},  $E_g\approx1.755$ eV, for a temperature of T=0 K.  Effective masses of charge carriers agree with those used in literature (see the next section). We should note that there are other publications with DFT results for the $\gamma$-CsPbI$_3$ crystal~\cite{Li2020, LinPSS2021}, however, they reproduce experimental data less accurately, in particular, the band gap.

The DFT calculations yield the model-accurate dispersion curves, which, however, are difficult to analyze. Therefore, we provide a simple explanation of the behavior of those curves based on the $\mathbf{k \cdot p}$ approach described, see, e.g., Ref.~\cite{yu_fundamentals_2010}. The $\mathbf{k \cdot p}$ method is based on the consideration of the local behavior of the dispersion dependences within the framework of perturbation theory. The coupling of the valence and conduction subbands is described within this approach by operators of the form:
\begin{equation}
V = \frac{\hbar}{2 m_0} k_\alpha p_{n,m}^{(\alpha)}. 
\label{eq:kp}
\end{equation}
Here $k_\alpha$ is the wave vector of an electron in the conduction band or a hole in the valence band, where $\alpha=x,y,z$ are the coordinates of the electron and hole within the unit cell of the crystal, and $m_0$ is the free electron mass. The matrix elements $p_{n,m}$ take the form, $p_{n,m}^{(\alpha)}=\langle u_{n}|\hat p_\alpha|u_{m}\rangle$, where the operator $\hat p_\alpha=-i\hbar\partial/\partial \alpha$ is the electron or hole momentum. The quantities $u_{n (m)}$ are the Bloch amplitudes of the electron or the hole wave functions, and indices $n,m$ determine the spin and orbital angular momentum of electrons and holes in conduction and valence bands, respectively.

The dispersion dependences in the vicinity of the $\Gamma$ point are of particular importance for the analysis of optical transitions near the band gap. The dispersion curves are mainly determined by the coupling of the topmost valence subband with the lowest conduction subband. This coupling is described by the matrix element of the operator (\ref{eq:kp}) and provides the main contribution to the effective masses of the electron and the hole near the $\Gamma$ point. The energies of electrons and holes can be calculated in the framework of the $\mathbf{k\cdot p}$ method using perturbation theory. This approach yields a parabolic dependence of the energies of both electrons and holes. This behavior is readily interpreted as the kinetic energy of free charge carriers with some effective mass $m^*$. 

The coupling of different subbands within the conduction or valence band can also contribute to the effective masses of the electron and hole. However, an analysis shows that this contribution is small for the nearest subbands in the vicinity of the $\Gamma$ point. The interaction of subbands becomes valuable for the critical points near the surface of the Brillouin zone. Such interactions contribute to the nonparabolicity of the dispersion dependences at large wave vectors.

\begin{figure}[t]
\centering
\includegraphics[width=\columnwidth]{"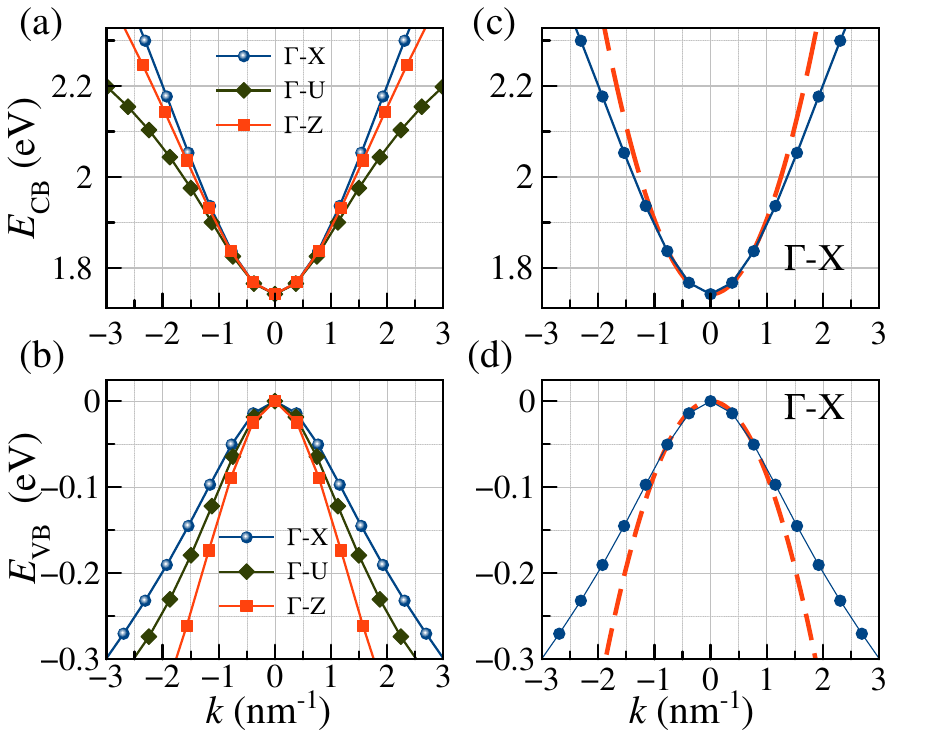"}
\caption{(a) and (b) The dispersion curves in the $\Gamma$-X, $\Gamma$-Z, and $\Gamma$-U directions for electrons and holes, respectively . (c) and (d) The effective mass approximation using Eq.~(\ref{eq:par0}) of the dispersion curves for the electrons and the holes, respectively, in the $\Gamma$-X direction. The points show the DFT calculation, the dashed lines show the parabolic approximation near the center of the Brillouin zone.}
\label{fig:DifferentDirections}
\end{figure}

The dispersion dependences for the $\Gamma$-X, $\Gamma$-Z, and $\Gamma$-U directions are shown in Fig.~\ref{fig:DifferentDirections}. The anisotropy (corrugation effect) and nonparabolicity of the curves are clearly seen here. The data obtained from the DFT calculations can be used to calculate the effective masses of the charge carriers. 

\section{The applicability of parabolic approximation}

Let us discuss the dispersion dependences using the effective mass model for the electron in the lower conduction subband and the hole in the upper valence subband, as well as estimate the energy range in which the model is applicable.

The dispersion dependences in a given direction in reciprocal space can be approximated, in the general case, by a polynomial of the wave vector. In semiconductors with a degenerate valence band, or in narrow-gap materials, the dispersion of charge carriers can be nonparabolic in the vicinity of the $\Gamma$-point \cite{Kane1957, AltarelliLipari1977, Loginov2025}. In lead-halide perovskites such as CsPbI$_3$, the band gap is wide enough, and both the lower conduction subband and the upper valence subband are only spin-degenerate. Therefore, for small enough values of $k$, we can use a single quadratic term of the polynomial that is the effective mass approximation
\begin{equation}
E (\vec{k}) =  \frac{\hbar^2 }{2 m^{*}_{\mathbf{j}}(0)} k^2.
\label{eq:par0}
\end{equation}
Here $E$ is the total energy of the quasiparticle, $k = |\vec{k}|$ is the absolute value of the wave vector, $m^{*}_{\mathbf{j}}(0)$ is the effective mass of the electron or the hole in the direction $\vec{k}$ in the parabolic approximation, and  $\mathbf{j} = \vec{k}/k$ is the unit vector. The expression for the inverse effective masses of the electron and hole in Eq. (\ref{eq:par0}) can be obtained within the $\mathbf{k \cdot p}$ method in the second-order perturbation theory of the operator given in Eq. (\ref{eq:kp}), see, e.g., Ref.~\cite{yu_fundamentals_2010}. As mentioned above, the values of the effective mass are determined by the coupling of the top valence subband and the bottom conduction subband.

The orthorhombic phase of CsPbI$_3$ exhibits significant anisotropy of dispersion dependencies, as demonstrated in Figs. \ref{fig:DifferentDirections} (a) and (b) for the directions $\Gamma$-X, $\Gamma$-Z, and $\Gamma$-U in the Brillouin zone. Figure \ref{fig:DifferentDirections} (c) and (d) show an example of the parabolic approximation of dispersion curves at small $k$ values. As seen, the approximation well describes the dispersion curve in a small energy range, ~0.1 eV for the valence band and ~0.2 eV for the conduction band.

To calculate the effective mass tensor values for each direction in the Brillouin zone, we approximate the points obtained from the DFT calculations. However, only a few of these points fall within the parabolic dispersion region. Therefore, the accuracy of the approximation for points within the parabolic region can be quite low. To obtain more accurate values of $m^*_e(0)$ and $m^*_h(0)$, we used a non-parabolic approximation, described in the next section. The results obtained from the non-parabolic approximation, which also uses $m^*_\alpha(0)$, are given in Table~\ref{tab:coefficients}.

The effective masses of the charge carriers obtained from the least-squares fitting of the dispersion curves in all the studied directions, as well as the average effective mass, are given in Tab.~\ref{tab:effective-masses}. The fit interval has been chosen as [-0.5, 0.5] nm$^{-1}$ for the $\Gamma$-Y and $\Gamma$-U directions and [-1, 1] nm$^{-1}$ for the other directions. The average effective mass of the electron, $\bar{m}^{*}_e(0)$, is given by
\begin{equation}
\frac{1}{\bar{m}^{*}_{e}(0)} = \frac{1}{N}\sum_{\mathbf{j}}^N \frac{1}{m^{*}_{e,\mathbf{j}}(0)}
\end{equation}
where the sum is carried out over all directions in the reciprocal space calculated by DFT, and $m^{*}_{e,\mathbf{j}}(0)$ is the effective mass of the electron in the corresponding direction. A similar expression should be used for the average effective mass of the hole. Using these values, we have also calculated the reduced exciton effective mass,
\begin{equation}
\frac{1}{\mu_{X,\mathbf{j}}(0)} = \frac{1}{m_{e,\mathbf{j}}^{*}(0)}+\frac{1}{m_{h,\mathbf{j}}^{*}(0)}.
\end{equation}
\begin{table}[H]
\centering
\caption{The effective masses of electrons $m_e^{*}(0)$ and holes $m_h^{*}(0)$ (in units of free electron mass, $m_0$) in different directions in reciprocal space. }
\label{tab:effective-masses}
\vspace{0.5\baselineskip}
\begin{tabular}{|l|c|c|c|c|c|c|c|c|}
\hline
        & $\Gamma\!-\!\mathrm{Z}$ & $\Gamma\!-\!\mathrm{T}$ & $\Gamma\!-\!\mathrm{Y}$ & $\Gamma\!-\!\mathrm{S}$ & $\Gamma\!-\!\mathrm{X}$ & $\Gamma\!-\!\mathrm{U}$ & $\Gamma\!-\!\mathrm{R}$ & Average \\ \hline
$m_h^{*}(0)$   & 0.23           & 0.26           & 0.17           & 0.37           & 0.45           & 0.33           & 0.33 & 0.27      \\ \hline
$m_e^{*}(0)$   & 0.25           & 0.24           & 0.15           & 0.23           & 0.23           & 0.29           & 0.25 & 0.23      \\ \hline
$\mu_X (0)$     & 0.12           & 0.12           & 0.08           & 0.14           & 0.15           & 0.15           & 0.14 & 0.13      \\ \hline
\end{tabular}
\end{table}

The values that have been obtained for the average effective masses are in good agreement with the results obtained for orthorhombic CsPbI$_3$ in \cite{Sutton2018_CsPbI3_effective_mass} using the GW method ($0.23~m_0$ for electrons and $0.24~m_0$ for holes). The value of the average reduced exciton effective mass is also in good agreement with that reported, e.g., in Ref.~\cite{Sutton2018_CsPbI3_effective_mass}. 

\section{The non-parabolic dispersion and the corrugation effect}

The parabolic approximation presented above is applicable when the bottom conduction subband and the top valence subband are non-degenerate and strongly separate from the other conduction and valence subbands, respectively. Accordingly, their interaction can be neglected at sufficiently small $\vec{k}$ values. As shown in Fig.~\ref{fig:DifferentDirections}(c, d), the parabolic approximation is no longer applicable at energies of about $0.1$ eV (in absolute value) for holes and $0.2$ eV for electrons, and a more complex model should be applied to describe the dispersion dependences. This means that the perturbation theory used in the $\mathbf{k \cdot p}$ method is no longer correct for large wave vectors. The reason for that is the relatively large kinetic energy of the charge carrier motion, which becomes comparable to the energy separation between the sub-bands in the conduction and valence bands (see Fig. \ref{fig:DispersionDependences}). In this case, the nonparabolicity of electron (hole) dispersion is due to the coupling between the lowest conduction (upper valence) subband and the other conduction (valence) subbands. 

Figure~\ref{fig:DispersionDependences} shows that, when the upper and lower conduction subbands approach the high-symmetry points of the Brillouin zone surface, they become degenerate. The upper valence subbands exhibit similar behavior. The convergence of the subbands is accompanied by an increase in the degree of nonparabolicity. This behavior can, in principle, be described using the $\mathbf{k \cdot p}$ method near the respective point of the Brillouin zone. However, in our case, more precise results are given by the DFT calculations.

Here, we propose a model that allows one to describe the dispersion dependences in a large range of wave vectors, up to the Brillouin zone boundary in most the cases. The model is based on a phenomenological formula for the kinetic energy of a quasiparticle with an effective mass, which is parametrically dependent on the wave vector,
\begin{equation}
E (\vec{k}) =    \frac{\hbar^2 k^2}{2 m^*(\vec{k})}   = \frac{\hbar^2}{2 S(\vec{k})} \sum_{\alpha} \frac{1}{m^*_{\alpha} (0) }  k_{\alpha}^2 
\label{eq:Non-Parabolic}
\end{equation}
where $\alpha=x,y,z$ and $m^{*}_{\alpha} (0)$ are the components of the effective mass tensor given in Eq. (\ref{eq:par0}) and Tab.~\ref{tab:effective-masses}. The $\vec{k}$-dependent parameter $S$, 
\begin{equation}
S (\vec{k}) = \left(1 + \sum_{\beta} \sum_{\gamma} B_{\beta \gamma} |k_{\beta} k_{\gamma}|\right),
\label{eq:monk}
\end{equation}
describes the nonparabolicity and the corrugation effect. Here the indices $\beta,\gamma=x,y,z$, and the summation is carried out over $\beta$ and $\gamma$. The quantities $B_{\beta \gamma}$ are the elements of a 3x3 symmetric matrix $\mathbb{B}$ of real-valued coefficients. The matrix $\mathbb{B}$ is an effective tensor that describes the strength of the interaction between the bottom conduction subband for electrons (or the top valence subband for holes) and the other electron and hole subbands, respectively. The absolute value $|k_{\beta} k_{\gamma}|$ is used here as the Hamiltonian must be invariant with respect to the coordinate inversion $k_{\beta} \mapsto -k_{\beta}$ as a result of symmetry. It should also be noted that the sum containing $B_{\beta \gamma}$ is the same for all values of $\alpha$. According to Eqs.~(\ref{eq:Non-Parabolic}, \ref{eq:monk}), the inverse effective mass in the arbitrary direction is calculated as
\begin{equation}
\frac{1}{m^*(\vec{k})}=\frac{1}{S(\vec{k})}\sum_\alpha \frac{1}{m^*_{\alpha}(0)} \frac{k^2_{\alpha}}{k^2}.
\label{eq:mass}
\end{equation}

\begin{figure}[t]
\includegraphics[width=\columnwidth]{"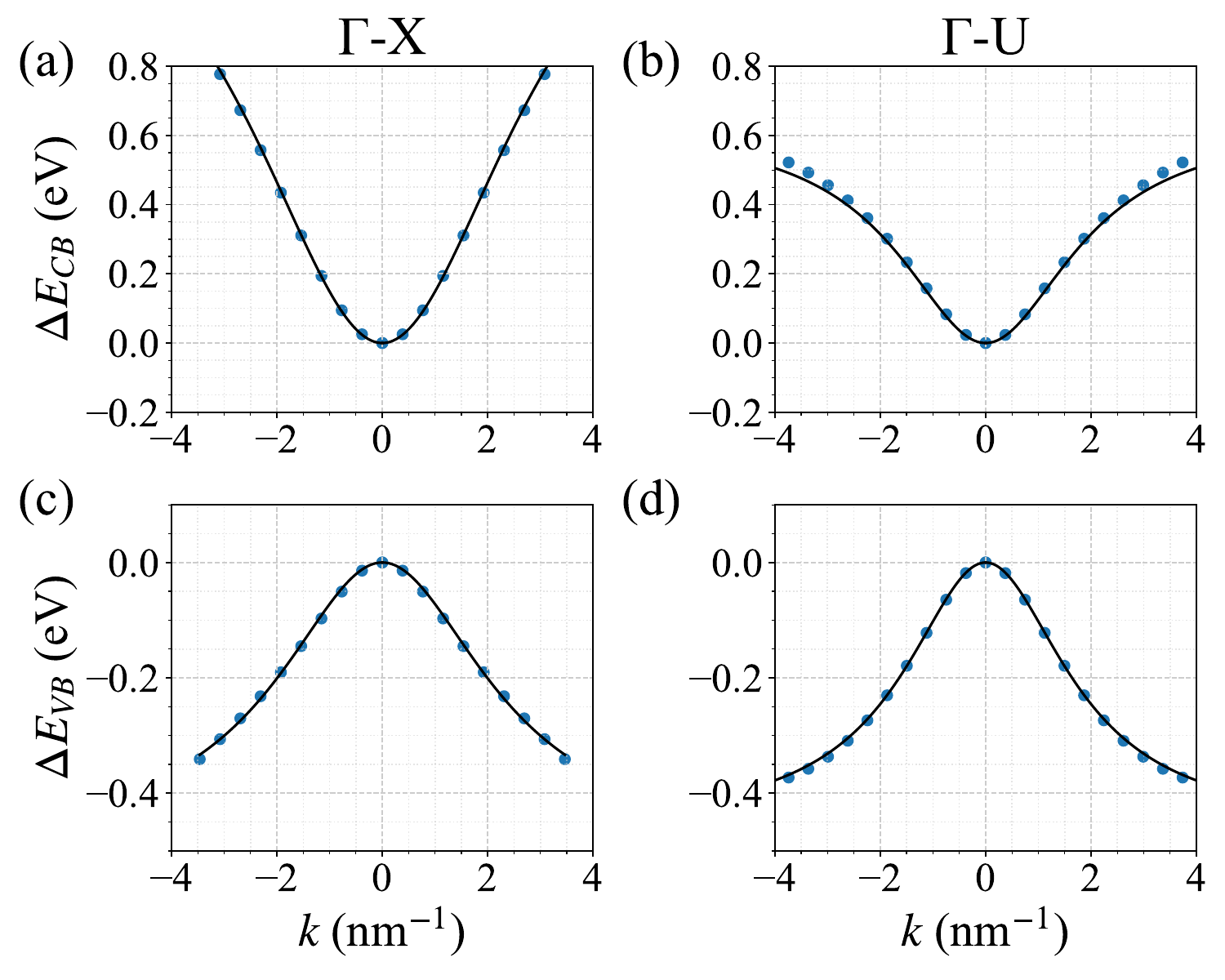"}
\caption{Examples of the dispersion curves for electrons (a,b) and holes (c,d) in directions $\Gamma$-X (a,c) and $\Gamma$-U (b,d) in the Brillouin zone, fitted using the nonparabolic model (see Eqs. (\ref{eq:Non-Parabolic})  and~(\ref{eq:monk})). The dots represent the DFT data, and the solid curves correspond to the least-squares fit of the data to the model. }
\label{fig:Approximation}
\end{figure}

As shown in Fig. \ref{fig:Approximation}, this approximation allows one to describe the dispersion curves with significantly better precision (the mean-square error is $4.3\cdot 10^{-4}$ eV for electrons and $6.5\cdot 10^{-5}$ eV for holes) in a larger energy range, up to $0.5$ eV above the bottom of the conduction band and up to $0.25$ eV below the top of the valence band. The coefficients of the approximation obtained for the different directions in reciprocal space are provided in Tab.~\ref{tab:coefficients}. The approximations for the other directions in the Brillouin zone are illustrated in Appendix~\ref{App:1}. It should be emphasized that only 9 model parameters were sufficient to describe the dispersion in the 7 directions of the Brillouin zone calculated by DFT.

We can also determine the effective mass for particular directions in the Brillouin zone. For example, for the $\Gamma$-U direction defined by the wave vector components $k_x = a_z k/\sqrt{a_x^2+a_z^2}$, $k_y = 0$, and $k_z = a_x k/\sqrt{a_x^2+a_z^2}$, Eq.~(\ref{eq:mass}) yields,
\begin{equation}
\begin{split}
E_U(\vec{k}) &= \frac{\hbar^2}{2(a_x^2+a_z^2)S(\vec{k})} \left(\frac{a_z^2}{m^*_x(0)} +\frac{a_x^2}{m^*_z(0)} \right)k^2 \\
&\equiv \frac{\hbar^2 k^2 }{2 m^*_{U}(\vec{k})},
\end{split}
\end{equation}
where
\begin{equation}
   S(\vec{k}) = 1 + \left[B_{xx}a^2_z+B_{zz}a^2_x+2B_{xz}a_xa_z\right]\frac{k^2}{a^2_x+a^2_z}.
\end{equation}
Here $a_x$ and $a_z$ are the lattice constants. It follows from this expression that the inverse effective mass in the $\Gamma-\mathrm{U}$ direction is calculated as
\begin{equation}
\frac{1}{m^*_{U}(\vec{k})} = \frac{1}{(a_x^2+a_z^2)S(\vec{k})} \left(\frac{a_z^2}{ m^*_x(0)} + \frac{a_x^2}{m^*_z(0)} \right).
\end{equation}
Similar expressions are valid for other directions, $\Gamma-\mathrm{S}$ and $\Gamma-\mathrm{T}$, by substituting $x \mapsto y$ and $z \mapsto y$, respectively. For the $\Gamma-\mathrm{R}$ direction, the expression for the inverse effective mass is more complex:
\begin{equation}
\begin{split}
\frac{1}{m^*_{R}(\vec{k})} &= \frac{1}{(a_x^2+a_y^2+a_z^2)S(\vec{k})} \\
&\times \left(\frac{a_y a_z}{ m^*_x(0)}+ \frac{a_x a_z}{m^*_y(0)} + \frac{a_x a_y}{m^*_z(0)} \right).
\end{split}
\end{equation}
The expression for $S(\vec{k})$ for this direction can be obtained from general expression~(\ref{eq:monk}).

\begin{table}[t]
	\caption{The nonparabolic dispersion coefficients (see Eq. (\ref{eq:Non-Parabolic})) for the electrons and the holes in reciprocal space obtained from the least-squares fit of the DFT data. Effective masses $m^*_{\alpha}(0)$ are given in units of free electron mass, and $B_{\beta \gamma}$ are given in nm$^2$.}
	\begin{center}
		\begin{tabular}{|c|c|c|c|c|}
			\hline
			Coefficient & \multicolumn{2}{c|}{Electrons} &\multicolumn{2}{c|}{Holes} \\
			\hline
			 & Value & Error & Value & Error \\
			\hline
			$m^*_x$ (0) & 0.233 & 0.8$\cdot 10^{-3}$ & 0.452 & 9.7$\cdot 10^{-3}$ \\
			\hline
			$m^*_y$ (0) & 0.150 & 2.0$\cdot 10^{-3}$ & 0.173 & 1.8$\cdot 10^{-3}$ \\
			\hline
			$m^*_z$ (0) & 0.254 & 3.1$\cdot 10^{-3}$ & 0.227 & 2.0$\cdot 10^{-3}$\\
			\hline
			$B_{xx}$ & 0.103 & 0.9$\cdot 10^{-5}$ & 0.170 & 1.4$\cdot 10^{-5}$  \\
			\hline
			$B_{yy}$ & 0.300 & 12.0$\cdot 10^{-5}$ & 0.639 & 14.0$\cdot 10^{-5}$ \\
			\hline
			$B_{zz}$ & 0.111 & 0.7$\cdot 10^{-5}$ & 0.235 & 0.6$\cdot 10^{-5}$ \\
			\hline
			$B_{xy}$ & -0.217 & 7.4$\cdot 10^{-5}$ & -0.211 & 7.3$\cdot 10^{-5}$ \\
			\hline
			$B_{yz}$ & -0.223 & 6.0$\cdot 10^{-5}$ & -0.347 & 5.9$\cdot 10^{-5}$ \\
			\hline
			$B_{zx}$ & 0.280 & 3.4$\cdot 10^{-5}$ & 0.159 & 3.0$\cdot 10^{-5}$ \\
			\hline
		\end{tabular}
	\end{center}
	\label{tab:coefficients}
\end{table}
% The phenomenological model based on Eqs. (\ref{eq:Non-Parabolic}) and (\ref{eq:monk}) can be useful for theoretical modeling of the quantum-confined states of electrons and holes in various nanocrystals, see, e.g., Refs.~\cite{Ekimov1996, Pages2000, Akopyan2005}. Usage of analytical expressions (\ref{eq:Non-Parabolic}) and (\ref{eq:monk}) for the Hamiltonian of the quantum problem, rather than the numerical DFT results, allows one to formulate and solve the problem in a traditional way. We will discuss the problem elsewhere~\cite{Sultanov_QCExciton_Unpublished}.  %Removed to not include unpublished work

\begin{figure}[t]
\centering
\includegraphics[width=\columnwidth]{"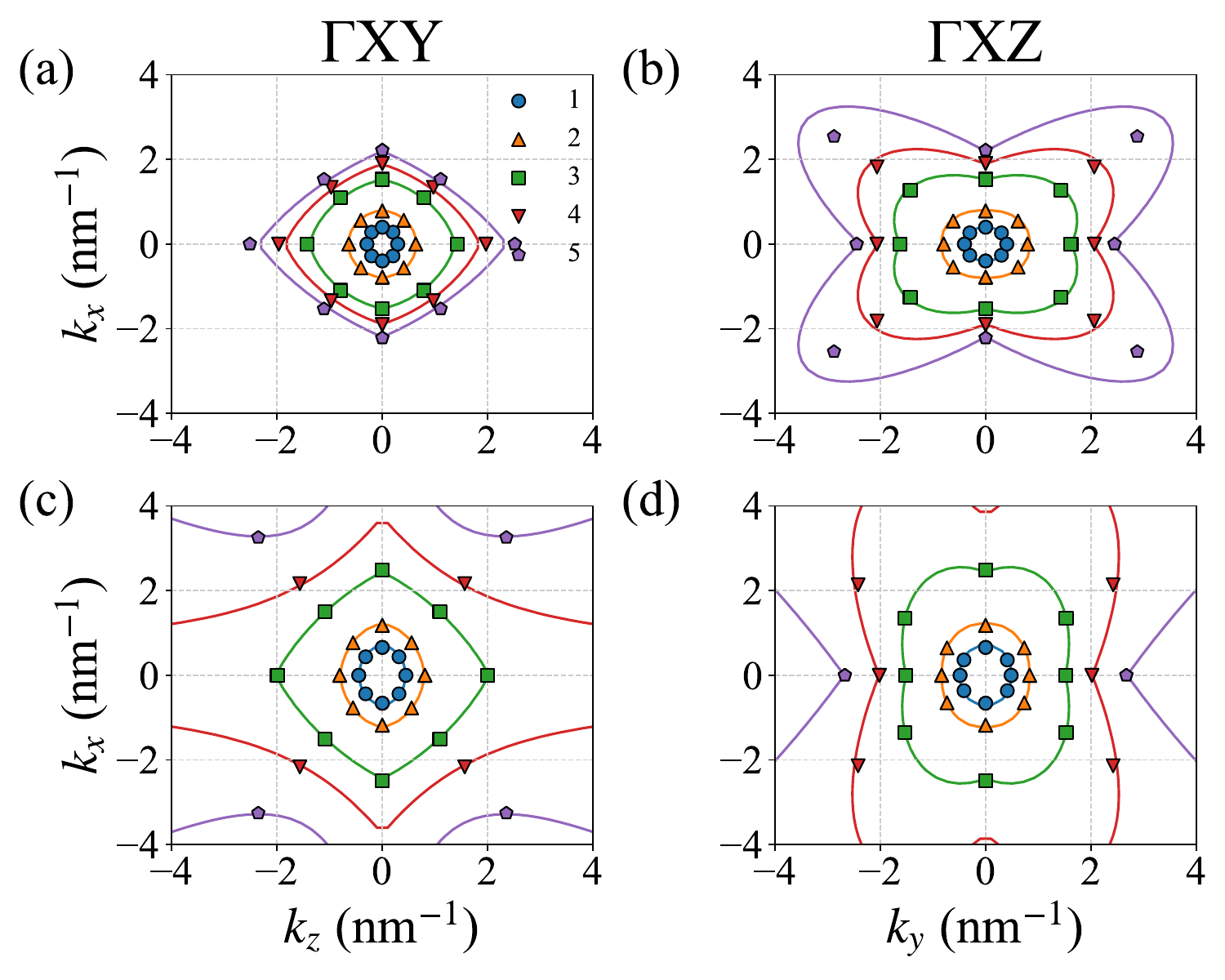"}
\caption{The cross-sections for energy isosurfaces for electron (top) and hole (bottom) in the $\Gamma$XZ (a,c) and the $\Gamma$XY (b,d) planes. Symbols are the DFT calculations. Solid lines are the modeling using Eqs.~(\ref{eq:Non-Parabolic}),~(\ref{eq:monk}), and values of parameters listed in Tab.~\ref{tab:coefficients}. The cross-section energies are, $\Delta E_{e,1}=0.03$ eV; $\Delta E_{e,2}=0.10$ eV; $\Delta E_{e,3}=0.31$ eV; $\Delta E_{e,4}=0.43$ eV; $\Delta E_{e,5}=0.53$ eV for electrons, and $\Delta E_{h,1}= -0.04$ eV; $\Delta E_{h,2}= -0.10$ eV; $\Delta E_{h,3}= -0.25$ eV; $\Delta E_{h,4}= -0.35$ eV; $\Delta E_{h,5}= -0.45$ eV for holes.}
\label{fig:isolines}
\end{figure}

In addition to strong nonparabolicity, the problem of the quantum-confined states is complicated by a strong corrugation effect, that is, the dependence of the dispersion curve parameters on the wave vector direction in the reciprocal space, as it is already illustrated in Fig. (\ref{fig:DifferentDirections}) and Tabs.~\ref{tab:effective-masses} and~\ref{tab:coefficients}. To visualize the corrugation effect more clearly, we present in Fig.~(\ref{fig:isolines}) the curves of constant energy in different planes in the reciprocal space for the charge carriers. These curves are the cross-section of the energy isosurfaces of the electron and the hole. 

As is evident, the curves of constant energy are circles or weakly elongated ellipses at small wave vectors and, respectively, at a small energy shift, $\Delta E$, from the bottom (top) of the conduction (valence) band. This is the area in $k$ space where the parabolic fit of the dispersion curves is a good approximation, see Fig.~\ref{fig:DifferentDirections}(b).

The curves of constant energy diverge from the elliptic-like shape at larger wave vectors and larger $\Delta E$. For example, the curve of constant energy has a more complex shape for $\Delta E_{e,5} = 0.53$~eV, see Fig.~\ref{fig:isolines} (c). In the valence band, they are significantly elongated along the $k_x$ direction. This corrugation of the conduction and valence bands should affect the energy states of carriers and, in particular, split the states that are degenerate with respect to the angular momentum.

\begin{figure}[t]
\centering
\includegraphics[width=0.9\columnwidth]{"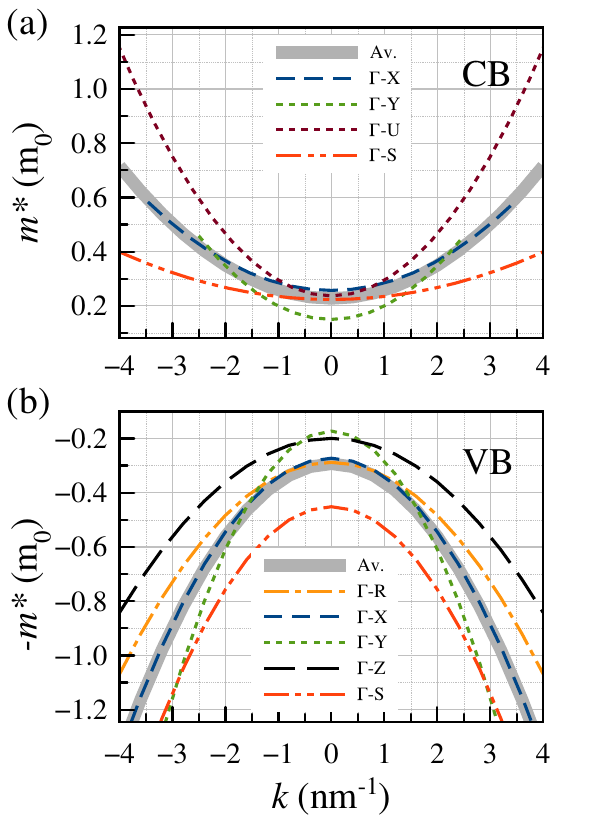"} 
\caption{Colored curves are the effective masses of electrons (a) and holes (b) in different directions in the Brillouin zone. Solid grey curves are the average effective mass derived using Eq.~(\ref{eq:average-integral}). The effective mass values for the hole are provided as negative values for posterity.}
\label{fig:effmasses}
\end{figure} 

For the solution of the quantum-confined state problem in the nanocrystals, whose size is considerably larger than the lattice constant, an envelope wave function approximation can be employed. In this case the effective mass of a charge carrier parametrically dependent on the carrier wave vector is also a good approximation. Using Eq.~(\ref{eq:monk}) and Tab.~\ref{tab:coefficients}, we have calculated the electron and hole masses as functions of wave vectors in different directions. They are shown in Fig.~\ref{fig:effmasses}. The anisotropy, as well as a significant increase in the effective mass with the increase of the wave vector, is evident.

In addition, we have calculated the averaged electron and hole effective masses as functions of wave vectors. They are frequently used for simplified calculations of the quantum-confined states \cite{kulebyakina_temperature-dependent_2024}. The averaged effective masses are useful for the calculation of the wave functions corresponding to the ground states of the electron and the hole having the highest possible symmetry. We can assume that the charge carrier cloud in the lowest energy state (the s-like state) averages the structure properties in different directions, thus diminishing the corrugation effect.

We introduce the average effective mass as
\begin{equation}
\frac{1}{\bar{m}^{*}} (k) = \frac{1}{4\pi}\int_{0}^{\pi} \sin \theta d \theta \int_{0}^{2\pi} d \varphi \frac{1}{m^{*} (\vec{k})}, 
\label{eq:average-integral}
\end{equation}
where $m^{*}(\vec{k})$ is the effective mass of the electron or the hole defined by Eq.~(\ref{eq:mass}), $\theta$ and $\varphi$ are the angles in spherical coordinates, 
\begin{equation}
\begin{cases}
k_x = k \sin \theta \cos \varphi \\
k_y = k \sin \theta \sin \varphi \\
k_z = k \cos \theta. 
\end{cases}
\end{equation}
The average effective mass introduced in this way can then be used to estimate the degree of anisotropy in the system.

As can be seen, the electron effective masses are greater than the average mass in the $\Gamma$-U direction. In both the $\Gamma$-X and $\Gamma$-Z directions, the electron effective masses are approximately the same as their average electron mass. In the $\Gamma$-T direction, the electron effective mass is smaller than the average one.

At the same time, the absolute value of the effective mass of the hole in the $\Gamma$-T direction is greater than its average hole mass. The absolute values of the hole masses in the $\Gamma$-U and $\Gamma$-Z directions are greater than the average value.

It should be noted that the averaged reduced mass can be used to analyze the energy of charge carriers in spherical nanocrystals of $\gamma$-CsPbI$_3$. The spherical shape of the nanocrystals imposes spherically symmetric boundary conditions on the anisotropic crystal structure of the perovskite. Therefore, the energy of the transition between the upper hole and the lower electron energy levels of size quantization can be described using the average inverse effective mass (Eq. (\ref{eq:average-integral})).

\section{Conclusion}

A comprehensive DFT calculation of the energy band structure in the $\gamma$-CsPbI$_3$ orthorhombic perovskite accounting for the spin-orbit interaction is performed. It is established that at sufficiently large wave vectors, the dispersion dependences of charge carriers in both the lower conduction subband and the upper valence subband are strongly nonparabolic. The effective mass approximation only holds for the absolute values of energy up to $0.1$ eV for holes and $0.2$ eV for electrons with respect to the edge of the band gap. Additionally, a significant corrugation effect has been observed.

A phenomenological model is proposed that describes the dispersion dependences of charge carriers in a large range of wave vectors. It is found that the dispersion in the 7 symmetric directions in the Brillouin zone can be described by the nine-parametric model. The effective mass is found to quadratically depend on the wave vector within this model. As such, the model can accurately describe the nonparabolicity of the dispersion, as well as the corrugation effect for the charge carriers in perovskites.

We believe that the proposed phenomenological model is not specific for the crystal under study and can be applied to the description of dispersion dependences of charged carriers in other crystals. The proposed model with the $k$-dependent effective mass is able to describe dispersion curves in a much larger wave vector range, up to the anticrossing point with other subbands, than the model with the constant effective mass. The model is useful for the analysis of optical spectra of the perovskite nanocrystals, which are small enough to observe the highly excited states of electrons and holes lying in the energy range of the substantial nonparabolicity of dispersion.

\section{Funding}
The authors acknowledge the Saint-Petersburg University for the Research Grant No. 125022803069-4.

\section{Acknowledgements}
The authors express their gratitude to the Computing Center of Research Park of Saint Petersburg State University.

\section{Author contributions}
Sultanov O.S - Conceptualization, Methodology, Formal Analysis, Visualization, Writing(Original Draft); Loginov D.K - Writing(Review and Editing); Ignatiev I.V - Supervision, Conceptualization, Writing(Review and Editing); Pankin D. V. - Software, Investigation; Smirnov M. B. - Conceptualization, Supervision; Kuznetsova M. S. - Conceptualization, Writing(Review and Editing).

\section{Conflicts of interest}
There are no conflicts to declare.

\section{Data availability}
The data used in the paper is available in the NOMAD repository under the DOI: 10.17172/NOMAD/2026.03.30-1 .

\onecolumngrid
\appendix
\section{The dispersion curves of charge carriers}
\label{App:1}
\begin{figure}[H] 
\centering
\includegraphics[width=\columnwidth]{"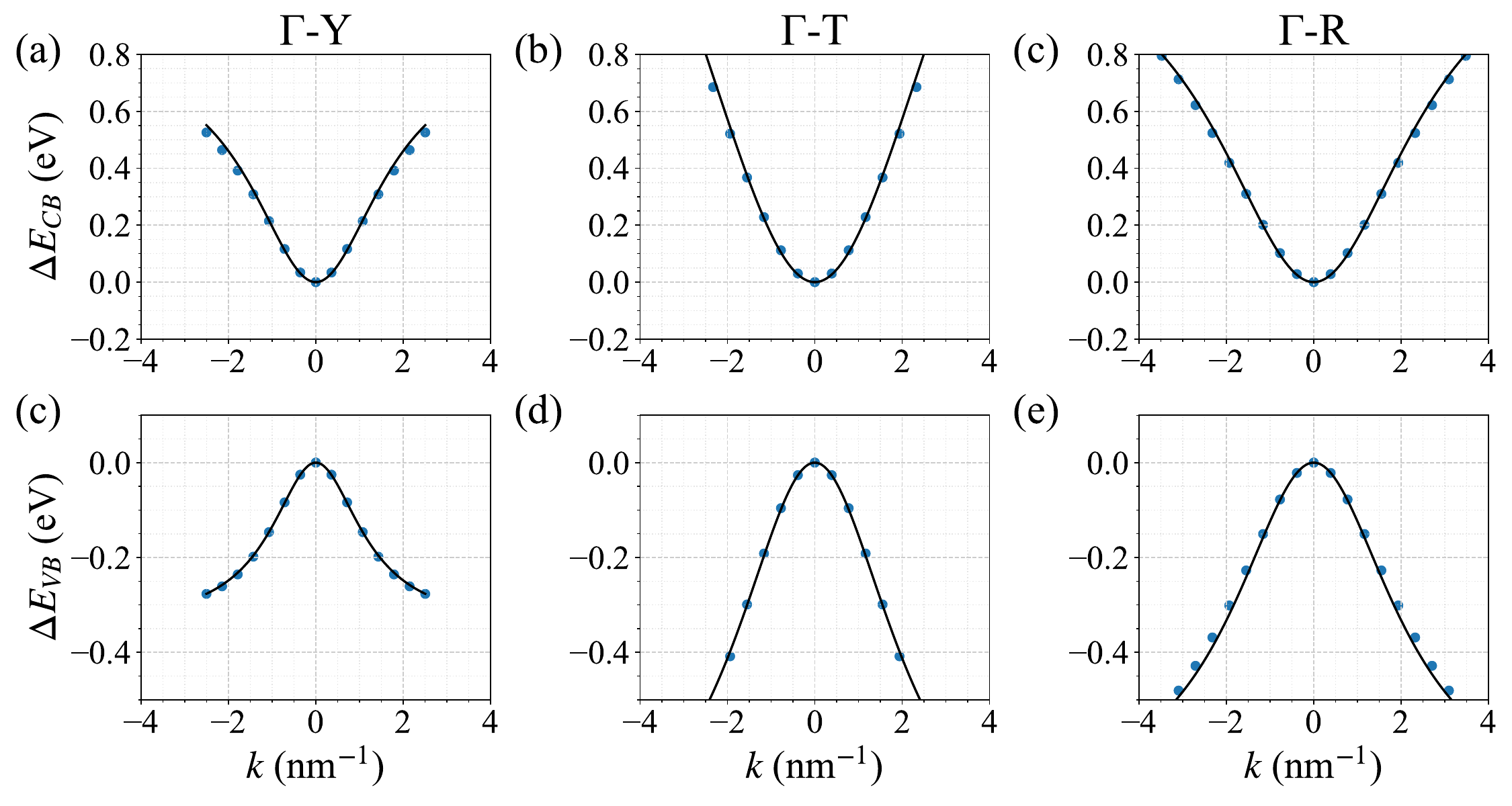"}
\caption{The dispersion curves of electrons (top) and holes (bottom) in the $\Gamma$-Y, $\Gamma$-T, and $\Gamma$-R directions, approximated by Eqs.~(\ref{eq:Non-Parabolic}) and~(\ref{eq:monk}) with parameters listed in Tab.~\ref{tab:coefficients}.}
\label{fig:Approximation2}
\end{figure}
\begin{figure}[H]
\begin{center}
\includegraphics[width=0.6\columnwidth]{"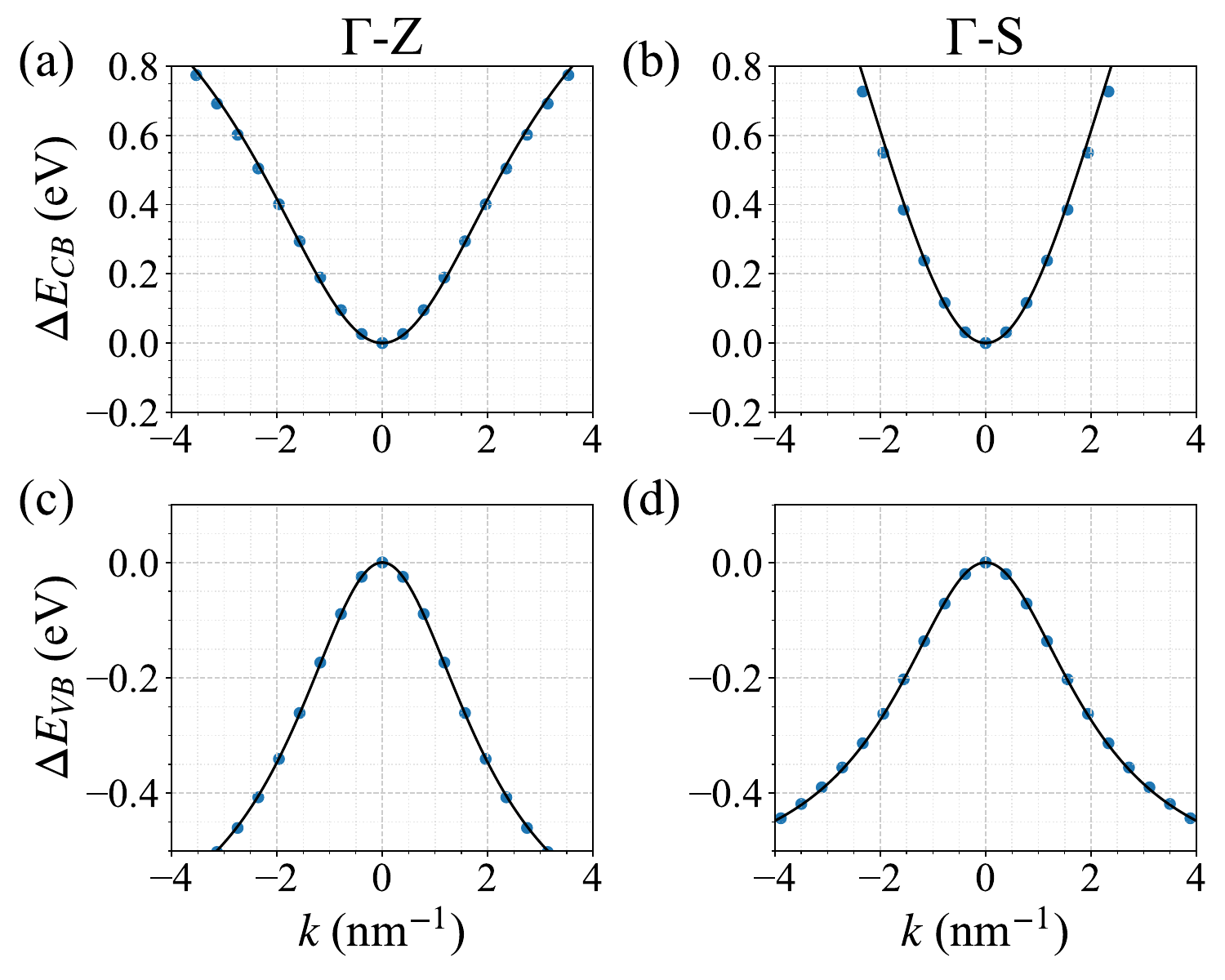"}
\caption{The dispersion curves of electrons (top) and holes (bottom) in the $\Gamma$-Z and $\Gamma$-S directions are approximated by Eqs.~(\ref{eq:Non-Parabolic}) and~(\ref{eq:monk}) with parameters listed in Tab.~\ref{tab:coefficients}.}
\label{fig:Approximation3}
\end{center}
\end{figure}
\bibliography{refs/refs} % Entries are in the refs folder

@article{tomic_influence_2004,
	title = {Influence of conduction-band nonparabolicity on electron confinement and effective mass in {GaN$_x$As$_{1-x}/$GaAs} quantum wells},
 	volume = {69},
 	url = {https://link.aps.org/doi/10.1103/PhysRevB.69.245305},
 	doi = {10.1103/physrevb.69.245305},
 	journal = {Phys. Rev. B},
 	author = {Tomić, Stanko and O’Reilly, Eoin P. and Klar, Peter J. and Grüning, Heiko and Heimbrodt, Wolfram and Chen, Weimin M. and Buyanova, Irina A.},
 	month = jun,
 	year = {2004},
}

@article{ekenberg_nonparabolicity_1989,
 	title = {Nonparabolicity effects in a quantum well: {Sublevel} shift, parallel mass, and {Landau} levels},
 	volume = {40},
 	url = {https://link.aps.org/doi/10.1103/PhysRevB.40.7714},
 	doi = {10.1103/physrevb.40.7714},
 	journal = {Phys. Rev. B},
 	author = {Ekenberg, U.},
 	month = oct,
 	year = {1989},
 	pages = {7714--7726},
}

@article{akkerman_genesis_2018,
	title = {Genesis, challenges and opportunities for colloidal lead halide perovskite nanocrystals},
 	volume = {17},
 	url = {https://www.nature.com/articles/s41563-018-0018-4},
 	doi = {10.1038/s41563-018-0018-4},
 	journal = {Nature Mater},
 	author = {Akkerman, Quinten A. and Rainò, Gabriele and Kovalenko, Maksym V. and Manna, Liberato},
 	month = may,
 	year = {2018},
 	pages = {394--405},
}

@article{nestoklon_tight-binding_2021,
	title = {Tight-binding description of inorganic lead halide perovskites in cubic phase},
 	volume = {196},
 	url = {https://linkinghub.elsevier.com/retrieve/pii/S0927025621002627},
 	doi = {10.1016/j.commatsci.2021.110535},
 	journal = {Computational Materials Science},
 	author = {Nestoklon, M.O.},
 	month = aug,
 	year = {2021},
 	pages = {110535},
}

@article{kirstein_mode_2023,
	title = {Mode locking of hole spin coherences in {CsPb}({Cl}, {Br})$_3$ perovskite nanocrystals},
	volume = {14},
	copyright = {2023 The Author(s)},
	issn = {2041-1723},
	url = {https://www.nature.com/articles/s41467-023-36165-0},
	doi = {10.1038/s41467-023-36165-0},
	language = {en},
	number = {1},
	urldate = {2025-07-24},
	journal = {Nat Commun},
	author = {Kirstein, E. and Kopteva, N. E. and Yakovlev, D. R. and Zhukov, E. A. and Kolobkova, E. V. and Kuznetsova, M. S. and Belykh, V. V. and Yugova, I. A. and Glazov, M. M. and Bayer, M. and Greilich, A.},
	month = feb,
	year = {2023},
	note = {Publisher: Nature Publishing Group},
	pages = {699},
}

@article{nestoklon_tailoring_2023,
	title = {Tailoring the {Electron} and {Hole} {Landé} {Factors} in {Lead} {Halide} {Perovskite} {Nanocrystals} by {Quantum} {Confinement} and {Halide} {Exchange}},
	volume = {23},
	copyright = {https://doi.org/10.15223/policy-029},
	issn = {1530-6984, 1530-6992},
	url = {https://pubs.acs.org/doi/10.1021/acs.nanolett.3c02349},
	doi = {10.1021/acs.nanolett.3c02349},
	language = {en},
	number = {17},
	urldate = {2025-07-24},
	journal = {Nano Lett.},
	author = {Nestoklon, Mikhail O. and Kirstein, Erik and Yakovlev, Dmitri R. and Zhukov, Evgeny A. and Glazov, Mikhail M. and Semina, Marina A. and Ivchenko, Eugeniyus L. and Kolobkova, Elena V. and Kuznetsova, Maria S. and Bayer, Manfred},
	month = sep,
	year = {2023},
	note = {Publisher: American Chemical Society (ACS)},
	pages = {8218--8224},
}

@article{kirstein_lande_2022,
	title = {The {Landé} factors of electrons and holes in lead halide perovskites: universal dependence on the band gap},
	volume = {13},
	copyright = {2022 The Author(s)},
	issn = {2041-1723},
	shorttitle = {The {Landé} factors of electrons and holes in lead halide perovskites},
	url = {https://www.nature.com/articles/s41467-022-30701-0},
	doi = {10.1038/s41467-022-30701-0},
	language = {en},
	number = {1},
	urldate = {2025-07-24},
	journal = {Nat Commun},
	author = {Kirstein, E. and Yakovlev, D. R. and Glazov, M. M. and Zhukov, E. A. and Kudlacik, D. and Kalitukha, I. V. and Sapega, V. F. and Dimitriev, G. S. and Semina, M. A. and Nestoklon, M. O. and Ivchenko, E. L. and Kopteva, N. E. and Dirin, D. N. and Nazarenko, O. and Kovalenko, M. V. and Baumann, A. and Höcker, J. and Dyakonov, V. and Bayer, M.},
	month = jun,
	year = {2022},
	note = {Publisher: Nature Publishing Group},
	pages = {3062},
}

@article{li_cspbx3_2016,
	title = {{CsPbX}$_{\textrm{3}}$ {Quantum} {Dots} for {Lighting} and {Displays}: {Room}‐{Temperature} {Synthesis}, {Photoluminescence} {Superiorities}, {Underlying} {Origins} and {White} {Light}‐{Emitting} {Diodes}},
	volume = {26},
	copyright = {http://onlinelibrary.wiley.com/termsAndConditions\#vor},
	issn = {1616-301X, 1616-3028},
	shorttitle = {{CsPbX}$_{\textrm{3}}$ {Quantum} {Dots} for {Lighting} and {Displays}},
	url = {https://onlinelibrary.wiley.com/doi/10.1002/adfm.201600109},
	doi = {10.1002/adfm.201600109},
	language = {en},
	number = {15},
	urldate = {2025-07-24},
	journal = {Adv Funct Materials},
	author = {Li, Xiaoming and Wu, Ye and Zhang, Shengli and Cai, Bo and Gu, Yu and Song, Jizhong and Zeng, Haibo},
	month = apr,
	year = {2016},
	note = {Publisher: Wiley},
	pages = {2435--2445},
}

@article{yakunin_low-threshold_2015,
	title = {Low-threshold amplified spontaneous emission and lasing from colloidal nanocrystals of caesium lead halide perovskites},
	volume = {6},
	copyright = {https://creativecommons.org/licenses/by/4.0},
	issn = {2041-1723},
	url = {https://www.nature.com/articles/ncomms9056},
	doi = {10.1038/ncomms9056},
	language = {en},
	number = {1},
	urldate = {2025-07-24},
	journal = {Nat Commun},
	author = {Yakunin, Sergii and Protesescu, Loredana and Krieg, Franziska and Bodnarchuk, Maryna I. and Nedelcu, Georgian and Humer, Markus and De Luca, Gabriele and Fiebig, Manfred and Heiss, Wolfgang and Kovalenko, Maksym V.},
	month = aug,
	year = {2015},
	note = {Publisher: Springer Science and Business Media LLC},
}

@article{ramasamy_all-inorganic_2016,
	title = {All-inorganic cesium lead halide perovskite nanocrystals for photodetector applications},
	volume = {52},
	issn = {1359-7345, 1364-548X},
	url = {https://xlink.rsc.org/?DOI=C5CC08643D},
	doi = {10.1039/c5cc08643d},
	language = {en},
	number = {10},
	urldate = {2025-07-24},
	journal = {Chem. Commun.},
	author = {Ramasamy, Parthiban and Lim, Da-Hye and Kim, Bumjin and Lee, Seung-Ho and Lee, Min-Sang and Lee, Jong-Soo},
	year = {2016},
	note = {Publisher: Royal Society of Chemistry (RSC)},
	pages = {2067--2070},
}

@article{eperon_formamidinium_2014,
	title = {Formamidinium lead trihalide: a broadly tunable perovskite for efficient planar heterojunction solar cells},
	volume = {7},
	issn = {1754-5692, 1754-5706},
	shorttitle = {Formamidinium lead trihalide},
	url = {https://xlink.rsc.org/?DOI=c3ee43822h},
	doi = {10.1039/c3ee43822h},
	language = {en},
	number = {3},
	urldate = {2025-07-24},
	journal = {Energy Environ. Sci.},
	author = {Eperon, Giles E. and Stranks, Samuel D. and Menelaou, Christopher and Johnston, Michael B. and Herz, Laura M. and Snaith, Henry J.},
	year = {2014},
	note = {Publisher: Royal Society of Chemistry (RSC)},
	pages = {982},
}

@article{saouma_selective_2017,
	title = {Selective enhancement of optical nonlinearity in two-dimensional organic-inorganic lead iodide perovskites},
	volume = {8},
	copyright = {https://creativecommons.org/licenses/by/4.0},
	issn = {2041-1723},
	url = {https://www.nature.com/articles/s41467-017-00788-x},
	doi = {10.1038/s41467-017-00788-x},
	language = {en},
	number = {1},
	urldate = {2025-07-24},
	journal = {Nat Commun},
	author = {Saouma, F. O. and Stoumpos, C. C. and Wong, J. and Kanatzidis, M. G. and Jang, J. I.},
	month = sep,
	year = {2017},
	note = {Publisher: Springer Science and Business Media LLC},
}

@article{wang_polarized_2016,
	title = {Polarized emission from {CsPbX}$_{\textrm{3}}$ perovskite quantum dots},
	volume = {8},
	issn = {2040-3364, 2040-3372},
	url = {https://xlink.rsc.org/?DOI=C6NR01915C},
	doi = {10.1039/c6nr01915c},
	language = {en},
	number = {22},
	urldate = {2025-07-24},
	journal = {Nanoscale},
	author = {Wang, Dan and Wu, Dan and Dong, Di and Chen, Wei and Hao, Junjie and Qin, Jing and Xu, Bing and Wang, Kai and Sun, Xiaowei},
	year = {2016},
	note = {Publisher: Royal Society of Chemistry (RSC)},
	pages = {11565--11570},
}

@article{pathak_perovskite_2015,
	title = {Perovskite {Crystals} for {Tunable} {White} {Light} {Emission}},
	volume = {27},
	issn = {0897-4756, 1520-5002},
	url = {https://pubs.acs.org/doi/10.1021/acs.chemmater.5b03769},
	doi = {10.1021/acs.chemmater.5b03769},
	language = {en},
	number = {23},
	urldate = {2025-07-24},
	journal = {Chem. Mater.},
	author = {Pathak, Sandeep and Sakai, Nobuya and Wisnivesky Rocca Rivarola, Florencia and Stranks, Samuel D. and Liu, Jiewei and Eperon, Giles E. and Ducati, Caterina and Wojciechowski, Konrad and Griffiths, James T. and Haghighirad, Amir Abbas and Pellaroque, Alba and Friend, Richard H. and Snaith, Henry J.},
	month = dec,
	year = {2015},
	note = {Publisher: American Chemical Society (ACS)},
	pages = {8066--8075},
}

@article{liu_halide-rich_2017,
	title = {Halide-{Rich} {Synthesized} {Cesium} {Lead} {Bromide} {Perovskite} {Nanocrystals} for {Light}-{Emitting} {Diodes} with {Improved} {Performance}},
	volume = {29},
	issn = {0897-4756, 1520-5002},
	url = {https://pubs.acs.org/doi/10.1021/acs.chemmater.7b00692},
	doi = {10.1021/acs.chemmater.7b00692},
	language = {en},
	number = {12},
	urldate = {2025-07-24},
	journal = {Chem. Mater.},
	author = {Liu, Peizhao and Chen, Wei and Wang, Weigao and Xu, Bing and Wu, Dan and Hao, Junjie and Cao, Wanyu and Fang, Fan and Li, Yang and Zeng, Yuanyuan and Pan, Ruikun and Chen, Shuming and Cao, Wanqiang and Sun, Xiao Wei and Wang, Kai},
	month = jun,
	year = {2017},
	note = {Publisher: American Chemical Society (ACS)},
	pages = {5168--5173},
}

@article{mei_hole-conductorfree_2014,
	title = {A hole-conductor–free, fully printable mesoscopic perovskite solar cell with high stability},
	volume = {345},
	issn = {0036-8075, 1095-9203},
	url = {https://www.science.org/doi/10.1126/science.1254763},
	doi = {10.1126/science.1254763},
	language = {en},
	number = {6194},
	urldate = {2025-07-24},
	journal = {Science},
	author = {Mei, Anyi and Li, Xiong and Liu, Linfeng and Ku, Zhiliang and Liu, Tongfa and Rong, Yaoguang and Xu, Mi and Hu, Min and Chen, Jiangzhao and Yang, Ying and Grätzel, Michael and Han, Hongwei},
	month = jul,
	year = {2014},
	note = {Publisher: American Association for the Advancement of Science (AAAS)},
	pages = {295--298},
}

@article{protesescu_nanocrystals_2015,
	title = {Nanocrystals of {Cesium} {Lead} {Halide} {Perovskites} ({CsPbX3}, {X} = {Cl}, {Br}, and {I}): {Novel} {Optoelectronic} {Materials} {Showing} {Bright} {Emission} with {Wide} {Color} {Gamut}},
	volume = {15},
	issn = {1530-6984},
	shorttitle = {Nanocrystals of {Cesium} {Lead} {Halide} {Perovskites} ({CsPbX3}, {X} = {Cl}, {Br}, and {I})},
	url = {https://doi.org/10.1021/nl5048779},
	doi = {10.1021/nl5048779},
	number = {6},
	urldate = {2025-07-24},
	journal = {Nano Lett.},
	author = {Protesescu, Loredana and Yakunin, Sergii and Bodnarchuk, Maryna I. and Krieg, Franziska and Caputo, Riccarda and Hendon, Christopher H. and Yang, Ruo Xi and Walsh, Aron and Kovalenko, Maksym V.},
	month = jun,
	year = {2015},
	note = {Publisher: American Chemical Society},
	pages = {3692--3696},
}

@article{yang_recent_2019,
	title = {Recent {Progress} on {Cesium} {Lead} {Halide} {Perovskites} for {Photodetection} {Applications}},
	volume = {1},
	copyright = {https://doi.org/10.15223/policy-029},
	issn = {2637-6113, 2637-6113},
	url = {https://pubs.acs.org/doi/10.1021/acsaelm.9b00346},
	doi = {10.1021/acsaelm.9b00346},
	language = {en},
	number = {8},
	urldate = {2025-07-24},
	journal = {ACS Appl. Electron. Mater.},
	author = {Yang, Tiebin and Li, Feng and Zheng, Rongkun},
	month = aug,
	year = {2019},
	pages = {1348--1366},
}

@article{kolobkova_perovskite_2021,
	title = {Perovskite {CsPbX3} ({X}={Cl}, {Br}, {I}) {Nanocrystals} in fluorophosphate glasses},
	volume = {563},
	copyright = {https://www.elsevier.com/tdm/userlicense/1.0/},
	issn = {0022-3093},
	url = {https://linkinghub.elsevier.com/retrieve/pii/S0022309321001708},
	doi = {10.1016/j.jnoncrysol.2021.120811},
	language = {en},
	urldate = {2025-07-24},
	journal = {Journal of Non-Crystalline Solids},
	author = {Kolobkova, E.V. and Kuznetsova, M.S. and Nikonorov, N.V.},
	month = jul,
	year = {2021},
	note = {Publisher: Elsevier BV},
	pages = {120811},
}

@article{pankin1,
	title = {First principles methods using {CASTEP}},
	volume = {220},
	issn = {2196-7105, 2194-4946},
	url = {https://www.degruyterbrill.com/document/doi/10.1524/zkri.220.5.567.65075/html},
	doi = {10.1524/zkri.220.5.567.65075},
	language = {en},
	number = {5-6},
	urldate = {2025-07-21},
	journal = {Zeitschrift für Kristallographie - Crystalline Materials},
	author = {Clark, Stewart J. and Segall, Matthew D. and Pickard, Chris J. and Hasnip, Phil J. and Probert, Matt I. J. and Refson, Keith and Payne, Mike C.},
	month = may,
	year = {2005},
	note = {Publisher: Walter de Gruyter GmbH},
	pages = {567--570},
}

@article{pankin2,
	title = {Accurate {Molecular} {Van} {Der} {Waals} {Interactions} from {Ground}-{State} {Electron} {Density} and {Free}-{Atom} {Reference} {Data}},
	volume = {102},
	copyright = {http://creativecommons.org/licenses/by/3.0/},
	issn = {0031-9007, 1079-7114},
	url = {https://link.aps.org/doi/10.1103/PhysRevLett.102.073005},
	doi = {10.1103/physrevlett.102.073005},
	language = {en},
	number = {7},
	urldate = {2025-07-21},
	journal = {Physical Review Letters},
	author = {Tkatchenko, Alexandre and Scheffler, Matthias},
	month = feb,
	year = {2009},
	note = {Publisher: American Physical Society (APS)},
}

@article{pankin3,
	title = {Variational density-functional perturbation theory for dielectrics and lattice dynamics},
	volume = {73},
	copyright = {http://link.aps.org/licenses/aps-default-license},
	issn = {1098-0121, 1550-235X},
	url = {https://link.aps.org/doi/10.1103/PhysRevB.73.155114},
	doi = {10.1103/physrevb.73.155114},
	language = {en},
	number = {15},
	urldate = {2025-07-21},
	journal = {Physical Review B},
	author = {Refson, Keith and Tulip, Paul R. and Clark, Stewart J.},
	month = apr,
	year = {2006},
	note = {Publisher: American Physical Society (APS)},
}

@misc{pankin4,
	title = {A {New} {CASTEP} and {ONETEP} {Geometry} {Optimiser}.},
	url = {http://www.hector.ac.uk/cse/distributedcse/reports/castep-geom/castep-geom/HTML/dCSE_project.html},
	urldate = {2025-07-15},
	author = {Aarons, J.},
}

@article{pankin5,
	title = {Representations of quasi-{Newton} matrices and their use in limited memory methods},
	volume = {63},
	issn = {0025-5610},
	number = {1-3},
	journal = {Math. Program.},
	author = {Byrd, Richard H. and Nocedal, Jorge and Schnabel, Robert B.},
	month = jan,
	year = {1994},
	pages = {129--156},
}

@article{pankin6,
	title = {Special points for {Brillouin}-zone integrations},
	volume = {13},
	copyright = {http://link.aps.org/licenses/aps-default-license},
	issn = {0556-2805},
	url = {https://link.aps.org/doi/10.1103/PhysRevB.13.5188},
	doi = {10.1103/physrevb.13.5188},
	language = {en},
	number = {12},
	urldate = {2025-07-21},
	journal = {Physical Review B},
	author = {Monkhorst, Hendrik J. and Pack, James D.},
	month = jun,
	year = {1976},
	note = {Publisher: American Physical Society (APS)},
	pages = {5188--5192},
}

@article{kulebyakina_temperature-dependent_2024,
	title = {Temperature-dependent photoluminescence dynamics of {Cs}{Pb}{Br}3 and {Cs}{Pb}({Cl},{Br})3 perovskite nanocrystals in a glass matrix},
	volume = {109},
	doi = {10.1103/PhysRevB.109.235301},
	number = {23},
	journal = {Phys. Rev. B},
	author = {Kulebyakina, E. V.},
	year = {2024},
}

@article{kovalenko_properties_2017,
	title = {Properties and potential optoelectronic applications of lead halide perovskite nanocrystals},
	volume = {358},
	url = {https://www.science.org/doi/10.1126/science.aam7093},
	doi = {10.1126/science.aam7093},
	journal = {Science},
	author = {Kovalenko, Maksym V. and Protesescu, Loredana and Bodnarchuk, Maryna I.},
	month = nov,
	year = {2017},
	pages = {745--750},
}

@article{dey_state_2021,
	title = {State of the {Art} and {Prospects} for {Halide} {Perovskite} {Nanocrystals}},
	volume = {15},
	url = {https://pubs.acs.org/doi/10.1021/acsnano.0c08903},
	doi = {10.1021/acsnano.0c08903},
	journal = {ACS Nano},
	author = {Dey, Amrita and Ye, Junzhi and De, Apurba and others},
	month = jul,
	year = {2021},
	pages = {10775--10981},
}

@article{nedelcu_fast_2015,
	title = {Fast {Anion}-{Exchange} in {Highly} {Luminescent} {Nanocrystals} of {Cesium} {Lead} {Halide} {Perovskites} ({CsPbX}$_{\textrm{3}}$, {X} = {Cl}, {Br}, {I})},
	volume = {15},
	url = {https://pubs.acs.org/doi/10.1021/acs.nanolett.5b02404},
	doi = {10.1021/acs.nanolett.5b02404},
	journal = {Nano Lett.},
	author = {Nedelcu, Georgian and Protesescu, Loredana and Yakunin, Sergii and Bodnarchuk, Maryna I. and Grotevent, Matthias J. and Kovalenko, Maksym V.},
	month = aug,
	year = {2015},
	pages = {5635--5640},
}

@article{diroll_hightemperature_2017,
	title = {High‐{Temperature} {Photoluminescence} of {CsPbX}$_{\textrm{3}}$ ({X} = {Cl}, {Br}, {I}) {Nanocrystals}},
	volume = {27},
	url = {https://onlinelibrary.wiley.com/doi/10.1002/adfm.201606750},
	doi = {10.1002/adfm.201606750},
	journal = {Adv Funct Materials},
	author = {Diroll, Benjamin T. and Nedelcu, Georgian and Kovalenko, Maksym V. and Schaller, Richard D.},
	month = jun,
	year = {2017},
}

@article{chen_twodimensional_2017,
	title = {Two‐{Dimensional} {Materials} for {Halide} {Perovskite}‐{Based} {Optoelectronic} {Devices}},
	volume = {29},
	url = {https://onlinelibrary.wiley.com/doi/10.1002/adma.201605448},
	doi = {10.1002/adma.201605448},
	journal = {Advanced Materials},
	author = {Chen, Shan and Shi, Gaoquan},
	month = jun,
	year = {2017},
}

@article{zhang_bication_2017,
	title = {Bication lead iodide {2D} perovskite component to stabilize inorganic $\alpha$-{CsPbI$_{\textrm{3}}$} perovskite phase for high-efficiency solar cells},
	volume = {3},
	url = {https://www.science.org/doi/10.1126/sciadv.1700841},
	doi = {10.1126/sciadv.1700841},
	journal = {Sci. Adv.},
	author = {Zhang, Taiyang and Dar, M. Ibrahim and Li, Ge and Xu, Feng and Guo, Nanjie and Grätzel, Michael and Zhao, Yixin},
	month = sep,
	year = {2017},
}

@article{zhang_stable_2024,
	title = {Stable $\alpha$-{CsPbI$_3$} with extremely red emission for expanding the color gamut},
	volume = {67},
	url = {https://link.springer.com/10.1007/s11432-023-3944-3},
	doi = {10.1007/s11432-023-3944-3},
	journal = {Sci. China Inf. Sci.},
	author = {Zhang, Yong and Wei, Xin and Gao, Lei and Zhao, Weijie and Sun, Changjiu and Wei, Junli and Yuan, Mingjian and Ni, Zhenhua and Lu, Junpeng and Liu, Hongwei},
	month = may,
	year = {2024},
}

@article{huang_exploration_2018,
	title = {The {Exploration} of {Carrier} {Behavior} in the {Inverted} {Mixed} {Perovskite} {Single}‐{Crystal} {Solar} {Cells}},
	volume = {5},
	url = {https://onlinelibrary.wiley.com/doi/10.1002/admi.201800224},
	doi = {10.1002/admi.201800224},
	journal = {Adv Materials Inter},
	author = {Huang, Yuan and Zhang, Yu and Sun, Junlu and Wang, Xiaoge and Sun, Junliang and Chen, Qi and Pan, Caofeng and Zhou, Huanping},
	month = jul,
	year = {2018},
}

@article{hodes_perovskite-based_2013,
	title = {Perovskite-{Based} {Solar} {Cells}},
	volume = {342},
	url = {https://www.science.org/doi/10.1126/science.1245473},
	doi = {10.1126/science.1245473},
	journal = {Science},
	author = {Hodes, Gary},
	month = oct,
	year = {2013},
	pages = {317--318},
}

@article{nie_high-efficiency_2015,
	title = {High-efficiency solution-processed perovskite solar cells with millimeter-scale grains},
	volume = {347},
	url = {https://www.science.org/doi/10.1126/science.aaa0472},
	doi = {10.1126/science.aaa0472},
	journal = {Science},
	author = {Nie, Wanyi and Tsai, Hsinhan and Asadpour, Reza and Blancon, Jean-Christophe and Neukirch, Amanda J. and Gupta, Gautam and Crochet, Jared J. and Chhowalla, Manish and Tretiak, Sergei and Alam, Muhammad A. and Wang, Hsing-Lin and Mohite, Aditya D.},
	month = jan,
	year = {2015},
	pages = {522--525},
}

@article{cho_overcoming_2015,
	title = {Overcoming the electroluminescence efficiency limitations of perovskite light-emitting diodes},
	volume = {350},
	url = {https://www.science.org/doi/10.1126/science.aad1818},
	doi = {10.1126/science.aad1818},
	journal = {Science},
	author = {Cho, Himchan and Jeong, Su-Hun and Park, Min-Ho and Kim, Young-Hoon and Wolf, Christoph and Lee, Chang-Lyoul and Heo, Jin Hyuck and Sadhanala, Aditya and Myoung, NoSoung and Yoo, Seunghyup and Im, Sang Hyuk and Friend, Richard H. and Lee, Tae-Woo},
	month = dec,
	year = {2015},
	pages = {1222--1225},
}

@article{tan_bright_2014,
	title = {Bright light-emitting diodes based on organometal halide perovskite},
	volume = {9},
	url = {https://www.nature.com/articles/nnano.2014.149},
	doi = {10.1038/nnano.2014.149},
	journal = {Nature Nanotech},
	author = {Tan, Zhi-Kuang and Moghaddam, Reza Saberi and Lai, May Ling and Docampo, Pablo and Higler, Ruben and Deschler, Felix and Price, Michael and Sadhanala, Aditya and Pazos, Luis M. and Credgington, Dan and Hanusch, Fabian and Bein, Thomas and Snaith, Henry J. and Friend, Richard H.},
	month = sep,
	year = {2014},
	pages = {687--692},
}

@article{Sutton2018_CsPbI3_effective_mass,
  title        = {Cubic or Orthorhombic? Revealing the Crystal Structure of Metastable Black-Phase CsPbI$_3$ by Theory and Experiment},
  author       = {Sutton, R. and others},
  journal      = {Univ. of Oxford (preprint) / published work},
  year         = {2018},
  note         = {GW and LDA calculations, Table 2 reports effective masses for γ-CsPbI$_3$},
  url          = {https://ora.ox.ac.uk/objects/uuid\%3A9f2752a8-e949-4c1c-87a8-f1165be4f989/files/m0207002b602a909a675ae50175cd2512}
}

@article{AltarelliLipari1977,
  author  = {Altarelli, M. and Lipari, N. O.},
  title   = {Exciton dispersion in semiconductors with degenerate bands},
  journal = {Physical Review B},
  year    = {1977},
  volume  = {15},
  pages   = {4898--4906},
  doi     = {10.1103/PhysRevB.15.4898},
  url     = {https://doi.org/10.1103/PhysRevB.15.4898}
}

@article{Loginov2025,
  author  = {Loginov, D. K.},
  title   = {Exact numeric calculation of nonparabolicity of exciton dispersion in semiconductors with degenerate valence band},
  journal = {Physica B: Condensed Matter},
  year    = {2025},
  volume  = {715},
  pages   = {417534},
  doi     = {10.1016/j.physb.2025.417534},
  url     = {https://www.sciencedirect.com/journal/physica-b-condensed-matter/vol/715/suppl/C}
}

@article{Kane1957,
  author  = {Kane, E. O.},
  title   = {Band structure of indium antimonide},
  journal = {Journal of Physics and Chemistry of Solids},
  year    = {1957},
  volume  = {1},
  pages   = {249--261},
  note    = {originally published 1957},
  url     = {https://www.sciencedirect.com/science/article/pii/0022369757900136}
}

@article{kojima_organometal_2009,
	title = {Organometal {Halide} {Perovskites} as {Visible}-{Light} {Sensitizers} for {Photovoltaic} {Cells}},
	volume = {131},
	issn = {0002-7863, 1520-5126},
	url = {https://pubs.acs.org/doi/10.1021/ja809598r},
	doi = {10.1021/ja809598r},
	language = {en},
	number = {17},
	urldate = {2025-12-25},
	journal = {J. Am. Chem. Soc.},
	author = {Kojima, Akihiro and Teshima, Kenjiro and Shirai, Yasuo and Miyasaka, Tsutomu},
	month = may,
	year = {2009},
	pages = {6050--6051},
}

@article{lee_efficient_2012,
	title = {Efficient {Hybrid} {Solar} {Cells} {Based} on {Meso}-{Superstructured} {Organometal} {Halide} {Perovskites}},
	volume = {338},
	issn = {0036-8075, 1095-9203},
	url = {https://www.science.org/doi/10.1126/science.1228604},
	doi = {10.1126/science.1228604},
	language = {en},
	number = {6107},
	urldate = {2025-12-25},
	journal = {Science},
	author = {Lee, Michael M. and Teuscher, Joël and Miyasaka, Tsutomu and Murakami, Takurou N. and Snaith, Henry J.},
	month = nov,
	year = {2012},
	pages = {643--647},
}

@article{burschka_sequential_2013,
	title = {Sequential deposition as a route to high-performance perovskite-sensitized solar cells},
	volume = {499},
	copyright = {http://www.springer.com/tdm},
	issn = {0028-0836, 1476-4687},
	url = {https://www.nature.com/articles/nature12340},
	doi = {10.1038/nature12340},
	language = {en},
	number = {7458},
	urldate = {2025-12-25},
	journal = {Nature},
	author = {Burschka, Julian and Pellet, Norman and Moon, Soo-Jin and Humphry-Baker, Robin and Gao, Peng and Nazeeruddin, Mohammad K. and Grätzel, Michael},
	month = jul,
	year = {2013},
	pages = {316--319},
}

@book{yu_fundamentals_2010,
	address = {Berlin, Heidelberg},
	series = {Graduate {Texts} in {Physics}},
	title = {Fundamentals of {Semiconductors}: {Physics} and {Materials} {Properties}},
	copyright = {https://www.springernature.com/gp/researchers/text-and-data-mining},
	isbn = {978-3-642-00709-5 978-3-642-00710-1},
	shorttitle = {Fundamentals of {Semiconductors}},
	url = {https://link.springer.com/10.1007/978-3-642-00710-1},
	language = {en},
	urldate = {2025-12-26},
	publisher = {Springer Berlin Heidelberg},
	author = {Yu, Peter Y. and Cardona, Manuel},
	year = {2010},
	doi = {10.1007/978-3-642-00710-1},
}

@article{PhysRevB.99.085207,
  title = {Impact of nonparabolic electronic band structure on the optical and transport properties of photovoltaic materials},
  author = {Whalley, Lucy D. and Frost, Jarvist M. and Morgan, Benjamin J. and Walsh, Aron},
  journal = {Phys. Rev. B},
  volume = {99},
  issue = {8},
  pages = {085207},
  numpages = {11},
  year = {2019},
  month = {Feb},
  publisher = {American Physical Society},
  doi = {10.1103/PhysRevB.99.085207},
  url = {https://link.aps.org/doi/10.1103/PhysRevB.99.085207}
}

@article{horton_accelerated_2025,
	title = {Accelerated data-driven materials science with the {Materials} {Project}},
	volume = {24},
	issn = {1476-1122, 1476-4660},
	url = {https://www.nature.com/articles/s41563-025-02272-0},
	doi = {10.1038/s41563-025-02272-0},
	language = {en},
	number = {10},
	urldate = {2026-02-13},
	journal = {Nat. Mater.},
	author = {Horton, Matthew K. and Huck, Patrick and Yang, Ruo Xi and Munro, Jason M. and Dwaraknath, Shyam and Ganose, Alex M. and Kingsbury, Ryan S. and Wen, Mingjian and Shen, Jimmy X. and Mathis, Tyler S. and Kaplan, Aaron D. and Berket, Karlo and Riebesell, Janosh and George, Janine and Rosen, Andrew S. and Spotte-Smith, Evan W. C. and McDermott, Matthew J. and Cohen, Orion A. and Dunn, Alex and Kuner, Matthew C. and Rignanese, Gian-Marco and Petretto, Guido and Waroquiers, David and Griffin, Sinead M. and Neaton, Jeffrey B. and Chrzan, Daryl C. and Asta, Mark and Hautier, Geoffroy and Cholia, Shreyas and Ceder, Gerbrand and Ong, Shyue Ping and Jain, Anubhav and Persson, Kristin A.},
	month = oct,
	year = {2025},
	pages = {1522--1532},
}

@article{jain_commentary_2013,
	title = {Commentary: {The} {Materials} {Project}: {A} materials genome approach to accelerating materials innovation},
	volume = {1},
	issn = {2166-532X},
	shorttitle = {Commentary},
	url = {https://pubs.aip.org/apm/article/1/1/011002/119685/Commentary-The-Materials-Project-A-materials},
	doi = {10.1063/1.4812323},
	language = {en},
	number = {1},
	urldate = {2026-02-13},
	journal = {APL Materials},
	author = {Jain, Anubhav and Ong, Shyue Ping and Hautier, Geoffroy and Chen, Wei and Richards, William Davidson and Dacek, Stephen and Cholia, Shreyas and Gunter, Dan and Skinner, David and Ceder, Gerbrand and Persson, Kristin A.},
	month = jul,
	year = {2013},
	pages = {011002},
}

@article{Gau_JLum2023,
  author  = {D. L. Gau and I. Galain and I. Aguiar and R. E. Marotti},
  title   = {Origin of photoluminescence and experimental determination of exciton binding energy, exciton--phonon interaction, and {Urbach} energy in $\gamma$-$\mathrm{CsPbI_3}$ nanoparticles},
  journal = {J. Lumin.},
  year    = {2023},
  volume  = {257},
  eid     = {119765},
  doi     = {10.1016/j.jlumin.2023.119765}
}

@article{Li2020,
  author  = {Z. Li and Y. Qin and L. Dong and K. Li and Y. Qiao and W. Li},
  title   = {Elastic and electronic origins of strain-stabilized photovoltaic $\gamma$-$\mathrm{CsPbI_3}$},
  journal = {Phys. Chem. Chem. Phys.},
  year    = {2020},
  volume  = {22},
  pages   = {12706--12712},
  doi     = {10.1039/D0CP01649G}
}

@article{LinPSS2021,
  author  = {Zhonghai Lin and Jiayi Lei and Pingjian Wang and Ling Xu and Xiaoxiao Zhang and Yunxin Kang and Mingyu Chen and Guangfen Wei},
  title   = {Effects of bromine substitution and vacancy defects on the structural and electronic properties of black orthorhombic $\mathrm{CsPbI_3}$ perovskite},
  journal = {Phys. Status Solidi RRL},
  year    = {2021},
  volume  = {15},
  eid     = {2100277},
  doi     = {10.1002/pssr.202100277}
}
\end{document}